%% file: Benhar.tex
\documentclass[onecolumn,sort&compress,numbers]{els-mrw} 

\usepackage{amsmath,amssymb,amsfonts,amsthm,makeidx,graphicx}
\usepackage{txfonts}
\usepackage{helvet}
\usepackage{slashed} 
\usepackage{wasysym}
\usepackage{isotope}

\usepackage{hyperref}
\usepackage{doi}
\hypersetup{
    colorlinks=true,    
    linkcolor=blue,     
    filecolor=magenta,  
    urlcolor=cyan,      
    citecolor=green,    
    pdftitle={Document Title},
    pdfpagemode=FullScreen,
}

\begin{document}


\chapter{Neutrino Cross Sections: Low Energy}\label{chap1}

\author[1]{Omar Benhar}%


\address[1]{\orgname{INFN}, \orgdiv{Sezione di Roma}, \orgaddress{Piazzale Aldo Moro, 2. I-00185 Roma, Italy}}

\articletag{Chapter Article tagline: update of previous edition, reprint.}

\maketitle

\begin{abstract}[Abstract]
	Low-energy neutrino interactions with isolated nucleons are accurately described by the effective theory 
	based on Fermi's groundbreaking description of neutron $\beta$-decay. On the other hand, the extension of this scheme 
	to the case of neutrino interactions with nuclear matter\textemdash the  understanding of which is critical for the description of
	a variety of astrophysical processes\textemdash involves non trivial difficulties, originating from the complexity of 
	nuclear structure and dynamics. This chapter provides a concise introduction to the formalism of nuclear 
	many-body theory, suitable to perform theoretical calculations of the nuclear matter response to neutrino interactions, as 
	well as a detailed analysis of the relevant reaction mechanisms. The neutrino mean free path in nuclear matter and its 
	implications for the description of astrophysical processes are also discussed.
\end{abstract}

\begin{keywords}
 	neutrino scattering \sep nuclear matter \sep nucleon-nucleon correlations \sep neutrino mean free path \sep neutron stars  
\end{keywords}


\section*{Key points}
\begin{itemize}
	\item This chapter aims at providing a concise and yet self-contained introduction to neutrino interactions with nuclear matter, 
	in the regime in which weak interactions are described by Fermi's effective theory, and non relativistic nucleons are the relevant hadronic degrees of freedom.  
	\item The effects of nuclear dynamics on the weak transition amplitudes and response functions will be  explained 
	in a systematic fashion, using the predictions of the non interacting Fermi gas model as a baseline. 
	\item The concepts of mean field, short- and long-range correlations will be introduced, and described  within a unified formalism based on nuclear many-body theory.
	\item The determination of the neutrino mean free path in nuclear matter and its astrophysical implications
	will be discussed.

\begin{glossary}[Nomenclature]
        \begin{tabular}{@{}lp{34pc}@{}}
                CBF & Correlated Basis Functions\\
                CC & Charged Current \\
                CHF & Correlated Hartee-Fock \\
                CTD & Correlated Tamm-Dancoff \\
                LRC & Long-Range Correlations \\
                MFA & Mean-Field Approximation \\
                MFP & Mean Free Path \\
                NC & Neutral Current \\
                SRC & Short-Range Correlations\\
        \end{tabular}
\end{glossary}

\end{itemize}


\section{Introduction}
\label{intro}

The interactions between neutrinos with energies in the tens of MeV range and isolated nucleons can be understood within 
the conceptual framework of the standard model of electroweak interactions, and described to remarkable accuracy using the formalism based on Fermi's effective theory of neutron $\beta$-decay; for a pedagogical introduction, see, 
e.g., {\color{black} Refs.~\cite{RQM,EW}}. Neutrino interactions with nuclear 
matter, on the other hand, are far from being described at fully quantitative level, and their study should rather be regarded as a largely 
independent endeavour, involving challenging issues associated with the microscopic treatment  of interacting many-nucleon systems.
Theoretical calculations of the neutrino-nucleus cross section\textemdash which plays a critical role in a many astrophysical processes, including supernova explosions, neutron star cooling and binary neutron star mergers\textemdash necessarily rely on simplified models of nuclear structure and dynamics, the accuracy of which depends significantly on the kinematics of the process under consideration. 

Within the mean-field approximation underlying the nuclear shell model, nucleons bound in nuclear matter are assumed to behave 
as independent particles moving in an average potential. This picture, however, while being adequate to describe many features of nuclear dynamics, fails to capture the effects of correlations among the nucleons, which have long been 
recognised to be important.\footnote{In their classic book, published in the 1950s, Blatt and Weisskopf warned the reader 
that "The limitation of any independent particle model lies in its inability to encompass the correlation between the positions and spins of the various particles in the system"~\cite{Blatt:1952ije}.}
 
Short-range correlations give rise to virtual scattering processes, leading to the excitation of the participating nucleons to high momentum 
states lying above the Fermi surface. The ensuing quenching of the occupation probability of the states belonging to the Fermi sea\textemdash unambiguously confirmed by theoretical and experimental studies of electron-nucleus scattering in the kinematic regime in which the beam 
particles interact primarily  with individual nucleons~\cite{RevModPhys.65.817}\textemdash has been shown to bring about a sizeable 
suppression of the weak transition amplitudes in nuclear matter, with respect to the predictions of the mean-field approximation~\citep{burrows1998,burrows1999}. 

Long-range correlations, leading to the excitation of collective modes, are
also known to be important in electron-nucleus scattering processes. They are, in fact, dominant
in the region of low momentum transfer, in which the space resolution of the incoming electron 
is large, compared to the average nucleon-nucleon separation distance in the target nucleus, and the interaction involves 
many nucleons. A discussion of the effects of long-range correlations on the nuclear responses to electromagnetic interactions 
can be found in, e.g., Refs.~\cite{Dellafiore:1984ht,Bauer_2005} and references therein.

The rest of this chapter is structured as follows. After Sect.~\ref{nuN}, describing in detail the derivation of  the neutrino-nucleon cross section in free space, Sect.~\ref{nuN:inmatter} introduces a formalism\textemdash based on non relativistic nuclear many-body theory\textemdash designed to describe neutrino interactions in nuclear matter. 
The response of this system to weak interactions and its relation to the neutrino cross section are also discussed in Sect.~\ref{nuN:inmatter}, which includes a brief overview of nuclear matter structure and dynamics, Sect.~\ref{nuclear:matter}.
The approximations commonly employed to carry out theoretical calculations of the weak transition amplitudes and the associated  
response functions are analyzed in Sect.~\ref{responses}, while the derivation of the neutrino mean free path from the nuclear matter cross section is outlined in Sect.~\ref{response:MFP}.
Finally, in the concluding section the results reported in this chapter are summarised, and put in the broader context 
of neutrino reactions relevant to neutron star astrophysics.


\section{Neutrino-nucleon interactions in free space}
\label{nuN}

\begin{figure}[h!]
	\centering
	\includegraphics[width=.55\textwidth]{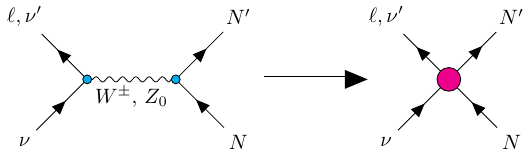}
{\color{black}	
	\caption{{\scriptsize Schematic representation of low-energy neutrino-nucleon scattering according to Fermi's effective 
	theory of weak interactions. The labels $\nu$, $\nu^\prime$ and $\ell$\textemdash referring both to the leptons and 
the corresponding antiparticles\textemdash denote the initial- and final-state neutrino and the associated 
	charged lepton, while $N$ and $N^\prime$ correspond to a neutron ($n$) or a proton ($p$).}}
}	
	\label{sec2:fig1}
		\end{figure}

According to the standard model of particle physics, neutrino interactions are mediated by the gauge bosons $W^{\pm}$ and
$Z_0$, the masses of which are $\sim$80 and $\sim$91 GeV, respectively. The tenet underlying Fermi's effective theory  
is that, owing to the large $W^{\pm}$ and $Z_0$ masses, weak interactions are short-ranged, and can be accurately approximated by a four-fermion contact interaction. Moreover, in the low-energy regime the internal structure  
of the nucleons does not come into play, and the neutrino-nucleon scattering process can be described as 
schematically illustrated by the diagrams of Fig.~\ref{sec2:fig1}. Note that $\nu$ and $\ell$ refer both to the neutral and charged leptons and to
the corresponding antiparticles. {\color{black} In this Chapter we will only consider reactions involving electron neutrinos and antineutrinos, and 
neglect the electron mass.} 

\subsection{Calculation of the scattering rate}
\label{W}

The Fermi Lagrangians describing charged-current (CC)  and neutral-current (NC) neutrino-nucleon  
interactions\textemdash associated with  $W^{\pm}$ and $Z_0$ exchange, respectively\textemdash
are written as a current-current product in the form
\begin{align}
\label{lagrangian}
\mathcal{L}_F(x) = \frac{G_w}{\sqrt{2}} \ {\ell}_\alpha(x) {j}^\alpha(x)   \ \ \ \ \ ,  \ {\rm with}  \ \ \ \ \  G_w = \left\{ 
\begin{array}{ll}
G_F & (NC) \\
G_F \cos \theta_C & (CC)
\end{array}
\right. \ . 
\end{align}
Here $G_F = 1.166 \times 10^{-5}  \ {\rm GeV}^{-2}$ and $\theta_C = 13.02^\circ$ denote the Fermi coupling constant and the Cabibbo mixing
angle, respectively, while the lepton and nucleon currents, $\ell_\alpha(x)$ and $j^\alpha(x)$, are given 
by\footnote{Throughout this chapter, $\psi_i$\textemdash  with the labels $i = \nu, \ell, n,  \ {\rm and} \ p$ 
corresponding to an electron neutrino, the associated charged lepton, a neutron and a proton\textemdash denotes a four-component Dirac field, and 
$\gamma^\alpha$ is a Dirac matrix.} 
\begin{align}
\label{currents:rel}
\ell_\alpha(x) = \left\{
\begin{array}{ll}
{\overline \psi}_{\ell}(x) \gamma_\alpha(1-\gamma_5)\psi_\nu(x) {\color{black} \ + \ h.c}.& (CC) \\
\overline{\psi}_{\nu^\prime}(x) \gamma_\alpha(1-\gamma_5)\psi_\nu(x) & (NC)
\end{array}
\right. \ \ \ \ \  ,  \ \ \ \ \ 
j^\alpha(x) = \left\{
\begin{array}{ll}
 \overline{\psi}_{p}(x) \gamma^\alpha(g_V- g_A \gamma_5)\psi_{n}(x) {\color{black} \ + \ h.c.} &  (CC) \\
   \overline{\psi}_{N^\prime}(x)  \frac{1}{2} \gamma^\alpha(c_V-c_A \gamma_5)\psi_{N}(x) & (NC)
\end{array}
\right. \  . 
\end{align}
{\color{black} where $g_V = 1$ and $g_A = 1.27$ are the vector and axial-vector CC neutrino-nucleon coupling constants, and $h.c.$ denotes the Hermitian Conjugate.
The corresponding neutral current couplings $c_V$ and $c_A$ are obtained from $g_A$ and $\sin^2 \theta_w = 0.231$, with $\theta_w$ being the weak mixing angle; see, e.g., Ref.~\cite{Horowitz}.}
At first order in the interaction Lagrangians defined by Eqs.\eqref{lagrangian} and \eqref{currents:rel}, the differential 
cross section of the  processes 
\begin{align}
\nu + n \to \ell + p \  \  \  (CC)  \  \  \  , \  \  \  \  \ \nu + N \to \nu^\prime + N^\prime \  \  \  (NC)  \ ,  
\end{align} 
whereby a neutrino with four-momentum $k\equiv(E,{\bf k})$ scatters off a nucleon {\color{black} $N$} of four-momentum $p\equiv(E_p,{\bf p})$
is given by
\begin{align}
\label{dsigma}
d\sigma_{\nu N} = W \frac{d^3 k^\prime}{(2 \pi)^3} \frac{d^3 p^\prime}{(2 \pi)^3} \ , 
\end{align}
where $k^\prime \equiv (E^\prime,{\bf k}^\prime)$ and  $p^\prime\equiv(E_{p^\prime},{\bf p}^\prime)$ denote the four-momenta of the final state lepton and nucleon, respectively. The scattering probability $W$ can be written as
\begin{align}
\label{def:W}
W = \frac{G_w^2}{2} \ \frac{L_{\alpha\beta} W^{\alpha\beta}}{2^4 E E^\prime E_{p} E_{p^\prime}} \  (2 \pi)^4 
\delta^{(4)}\left (k+p-k^\prime-p^\prime \right) \ ,
\end{align}
with the tensors $L_{\alpha\beta}$ and $W^{\alpha\beta}$ being defined in terms of transition matrix elements of the 
lepton and nucleon currents, respectively. Substituting 
the results reported in the box in the above equation one finds the expression 
\begin{align}
\nonumber
W = G_w^2 \ \frac{1}{E E^\prime E_p E_{p^\prime}} \Bigg\{ G_V^2 \ \Big[  (k p)(k^\prime p^\prime) & + (k p^\prime) (k^\prime p) 
- m_N^2 (k k^\prime) \Big] + 2 G_V G_A \Big[   (k p)(k^\prime p^\prime) - (k p^\prime)(k^\prime p)  \Big]  \\
\label{W:nuN}
& + G_A^2 \Big[ (k p)(k^\prime p^\prime) + (k p^\prime) (k^\prime p) + m_N^2 (k k^\prime)\Big]~\Bigg\}\ 
(2 \pi)^4\delta^{(4)} \left( k+p - k^\prime - p^\prime \right) \ , 
\end{align}
where $m_N$ denotes the nucleon mass\footnote{Here, and in what follows, the $\sim$1\permil~mass difference between the proton and the neutron will be neglected.} and
\begin{align}
\label{couplings}
G_V = \left\{
\begin{array}{ll}
g_V & (CC) \\
\frac{1}{2} c_V & (NC)
\end{array}
\right. \ \ \ \ \ , \ \ \ \ \ 
G_A = \left\{
\begin{array}{ll}
g_A  & (CC) \\
\frac{1}{2} c_A & (NC)
\end{array} \ .
\right. \ \ \ \ \
\end{align}

\begin{BoxTypeA}[W:box1]{Neutrino-nucleon scattering}

Consider, for the sake of simplicity,  a  NC neutrino-neutron interaction 
\begin{align}
\nonumber
\nu(k) + n(p) \to \nu^\prime(k^\prime) + n^\prime(p^\prime)
\end{align}
where $k \equiv (E,{\bf k})$, $k^\prime \equiv (E^\prime,{\bf k}^\prime)$, $p \equiv (E_p,{\bf p})$, and 
$p^\prime \equiv (E_{p^\prime},{\bf p}^\prime)$ denote the four-momenta of the participating particles. The expressions of the lepton 
and nucleon tensors appearing in the definition of the scattering probability, {\color{black} Eq.\eqref{def:W}, can be readily derived using standard 
properties of Dirac's $\gamma$-matrices and spinors; see, e.g., Refs.~\cite{RQM,EW}. The resulting expressions turn out to be}
\begin{align}
\label{Lab}
\nonumber
L_{\alpha\beta} & = \sum_{\rm spins}  {\overline u}_{\nu^\prime} \gamma_\alpha(1-\gamma_5)u_\nu  {\overline u}_{\nu} 
\gamma_\beta(1-\gamma_5)u_{\nu^\prime} = {\rm Tr}~\left[ \slashed{k}^\prime \gamma_\alpha(1-\gamma_5) \slashed{k} \gamma_\beta(1-\gamma_5) \right]~\\
& = 8 \Big[ k_\alpha k^\prime_\beta + k_\beta k^\prime_\alpha 
- g_{\alpha\beta}(k k^\prime)  - i \epsilon_{\alpha \beta \rho \sigma} {k^\prime}^\rho {k}^\sigma  \Big] \ , \\
\nonumber
W^{\alpha\beta} & = \frac{1}{2} \sum_{\rm spins}  {\overline u}_{n^\prime} \frac{1}{2} \gamma^\alpha (c_V -  c_A\gamma_5)u_n  {\overline u}_{n}
 \frac{1}{2} \gamma^\beta(c_V - c_A \gamma_5) u_{n^\prime}
 \nonumber
 = 
 \frac{1}{2} {\rm Tr}~\left[ (\slashed{p}^\prime +m_N) \frac{1}{2} \gamma^\alpha(c_V-c_A\gamma_5) (\slashed{p}+m_N)\frac{1}{2}
 \gamma^\beta(c_V-c_A\gamma_5) \right]~\\ 
  & = ( c_V^2 + c_A^2 )~\Big[ p^\alpha {p^\prime}^\beta + p^\beta {p^\prime}^\alpha
 - g^{\alpha\beta} (p p^\prime) \Big]  + ( c_V^2 - c_A^2 )~g^{\alpha\beta}m_N^2 - c_V c_A~i \epsilon^{\alpha \beta \rho \sigma} {p^\prime_\rho} {p}_\sigma  \  ,  
\label{Wab}
 \end{align}
where $u_i$,  with $i = \nu, \nu^\prime, n,  \ n,^\prime$, denotes a four-component Dirac spinor{\color{black} \textemdash normalised such that
${\overline u}_i u_i = 2 m_i$\textemdash}and $\epsilon_{\alpha \beta \rho \sigma}$ is the four-dimensional Levi-Civita tensor. 
Substitution of the above results  in Eq.~\eqref{def:W} yields the probability of NC neutrino-neutron interactions, $W$,  defined by 
Eqs.~\eqref{W:nuN} and~\eqref{couplings}.  
Note that, in the limit in which the masses of charged leptons can be neglected, the same expressions can be used to obtain 
the rate of CC interactions {\color{black} leading to neutron $\beta$-decay} by replacing  $c_V/2 \to g_V$ and $c_A/2 \to g_A$. 

\end{BoxTypeA}

\subsection{Non-relativistic limit}
\label{NONREL}

In the case of interactions with low-energy neutrinos, the non-relativistic treatment of nucleons is known to be very accurate. In this limit,    
the current $j^\alpha$ of Eq.~\eqref{currents:rel} can be written in the form
\begin{align}
\label{currents} 
j^\alpha \equiv (j^0,{\bf j})  \ \ , \ \  {\rm with} \ \left\{ 
\begin{array}{ll}
 j^0 = G_V~2m_N~\chi^\dagger_{s^\prime}~\chi_s = G_V~2m_N~\delta_{s s^\prime}  \\~\\
{\bf j}  = G_A~2m_N~\chi^\dagger_{s^\prime}~\boldsymbol{\sigma}~\chi_s \\
\end{array}
\right. \  , 
\end{align}
the coupling constants $G_V$ and $G_A$ being given by Eq.~\eqref{couplings}.
Here,  $\chi_s$ is a two-component Pauli spinor, $s$ and $s^\prime$ denote the projections of the nucleon spin in the initial and final state, respectively,  and ${\boldsymbol \sigma} \equiv(\sigma_1,\sigma_2,\sigma_3)$, with $\sigma_i$ being a Pauli matrix. {\color{black} Note that the above equation is obtained under the standard assumption that the momentum of the final state nucleon can be safely neglected.} 

The corresponding non-relativistic  expression of $W^{\alpha\beta}$ turns out to be diagonal, with 
$W^{00} = G_V^2~4 m_N^2$ and $W^{ii} = G_A^2~4 m_N^2$ for $i =$ 1, 2,  and 3. The scattering probability defined by Eq.~\eqref{def:W} reduces to  
\begin{align}
\label{W:N:NR}
W = G_w^2  \frac{1}{32 E E^\prime m_N^2}  \Big[ L_{00} W^{00} + L_{ii} W^{ii} \big] (2 \pi)^4\delta^{(4)} 
\left( k+{p} - k^\prime - {p}^\prime \right) = G_w^2 \Big[ G_V^2 (1 + \cos\theta) + G_A^2 ( 3 - \cos\theta) \Big]~(2\pi)^4\delta^{(4)}\left( k+{p} - k^\prime - {p}^\prime \right) \ ,
\end{align} 
where a sum over the index {\color{black} $i= 1,2,3$} is implied, and $\cos\theta = ( {\bf k} \cdot {\bf k}^\prime )/ ( | {\bf k}| |{\bf k}^\prime| $).
Note that the above expression can be also obtained from Eq.~\eqref{W:nuN}, by neglecting all contributions proportional to the 
nucleon velocities. 
 
 \section{Neutrino-nucleon interactions in nuclear matter}
 \label{nuN:inmatter}
 
The rate of neutrino-nucleon interactions in nuclear matter can be written in a form reminiscent of Eq.~\eqref{def:W} as
\begin{align}
\label{rate}
W =  \frac{G_w^2}{2} \frac{1}{ 4 E E^\prime} L_{\alpha \beta} S^{\alpha \beta} \ ,
\end{align}
with $L_{\alpha \beta}$ defined by Eq.\eqref{Lab}, and
\begin{align}
\label{nuclear:tensor}
S^{\alpha \beta}  = \sum_{F}~  \langle 0 | {J_A^\alpha}^\dagger | F  \rangle \langle F | J_A^\beta | 0  \rangle~(2 \pi)^4~\delta^{(4)}(P_0 + q  - P_F)  \ . 
\end{align}
In the above equation, $| 0 \rangle$ and $| F \rangle$ denote the nuclear matter ground and final states
{\color{black}  normalised such that $\langle 0 | 0 \rangle = \langle F | F \rangle = 1$}, having four-momentum
$P_0 \equiv  (E_0, {\bf P}_0)$ and  $P_F \equiv (E_F, {\bf P}_F)$, respectively, and $q = k - k^\prime \equiv(\omega, {\bf q})$ is the four-momentum 
transfer. The nuclear transition amplitude is given by 
\begin{align}
\label{nuclear:amplitude}
\mathcal{M}^\alpha = \langle F | J_A^\alpha | 0  \rangle  =  \langle F |~\sum_{i=1}^A j^\alpha_i~| 0  \rangle  \ ,  
\end{align}
{\color{black}
where $A$ denotes the number of target nucleons\footnote{In the nuclear matter limit both $A$ and the normalisation volume $V$ tend to infinity, 
with the nucleon density $\varrho = A/V$ remaining finite.} and the non-relativistic weak currents associated with the $i$-th nucleon can be conveniently written in the form $j^\alpha_i \equiv (j^0_i,{\bf j}_i)$, with 
\begin{align}
\label{nuclear:currents} 
j^0_i =  \ \left\{ 
\begin{array}{ll}
g_V~e^{ i {\bf q} \cdot {\bf r}_i } \tau^\pm &  (CC) \\
\frac{1}{2} c_V~e^{ i {\bf q} \cdot {\bf r}_i } &  (NC) \\
\end{array}
\right. \ \ \ \ \ , \ \ \ \ \  
{\bf j}_i =  \ \left\{ 
\begin{array}{ll}
g_A~e^{ i {\bf q} \cdot {\bf r}_i } \boldsymbol{\sigma}~\tau^\pm &  (CC) \\
\frac{1}{2} c_A~e^{ i {\bf q} \cdot {\bf r}_i } \boldsymbol{\sigma} &  (NC) \\
\end{array}
\right. \  . 
\end{align}
Within the formalism underlying the above equations, the nucleon is represented as an isospin doublet, the upper and lower components of 
which are the proton and neutron fields $\psi_p$ and $\psi_n$, while  $\tau^\pm = ( \tau_1 \pm i \tau_2)/2$\textemdash with the the $\tau_i$ 
being Pauli matrices acting in isospin space\textemdash denotes the isospin 
raising and lowering operators.

}

\subsection{Structure and dynamics of nuclear matter}
\label{nuclear:matter}

In uniform nuclear matter, translation invariance dictates that single-nucleon states must be eigenstates of 
the momentum operator labelled by the eigenvalue ${\bf k}$, with the  corresponding wave functions being  plane waves. 

The initial and final states appearing in Eq.~\eqref{nuclear:tensor} are solutions of the many-body Schr\"odinger equations 
\begin{align}
\label{schroedinger}
H |0\rangle = E_0 |0\rangle \ \ \ \ \ , \ \ \ \ \ H |F\rangle = E_F |F\rangle \ , 
\end{align}
where the nuclear Hamiltonian
\begin{align}
\label{Hamiltonian}
H = \sum_{i=1}^A \frac{{\bf k}_i^2}{2 m_N} + \sum_{j>i=1}^A v_{ij} + \sum_{k>j>i=1}^A V_{ijk} \ , 
\end{align}
is defined in terms of potentials describing nucleon-nucleon (NN) and {\color{black} {\em irreducible}} three-nucleon (NNN) interactions, providing  
an accurate description of the observed properties of the two- and three-nucleon systems.

Owing to the complexity of the NN and NNN interactions, $v_{ij}$ and $V_{ijk}$ comprise spin- and isospin-dependent as well as non-spherically symmetric terms. As a consequence, the solution of Eqs.~\eqref{schroedinger} for $A > 12$ involves prohibitive difficulties, and necessarily requires the use of suitable 
approximations schemes; a concise introduction to  the treatment of nuclear matter within 
many-body theory can be found in, e.g., Ref.~\cite{BF:NM}.

The independent particle approximation, underlying the nuclear shell model, is based on the tenet that the interaction terms in the right-hand side of Eq.~\eqref{Hamiltonian} can be replaced with a {\it mean field}. Therefore, this approach involves the substitution
\begin{align}
\sum_{j>i=1}^{\rm A} v_{ij} + \sum_{k>j>i=1}^{\rm A} V_{ijk} \rightarrow \sum_{i=1}^{\rm A} U_{i}\ ,
\label{meanfield}
\end{align}
with the potential $U_i$ chosen in such a way that the single-particle Hamiltonian
\begin{align}
 h_i = \frac{{\bf k}_i^2}{2m_N} + U_{k_i} \ ,
\label{sp:h}
\end{align}
is diagonal in momentum space. The ground-state state $|0\rangle$ is represented by the 
antisymmetric product of the $A$ lowest energy eigenstates of the operator $h_i$ corresponding to momenta such that  
$|{\bf k}_i| \leq k_F$. This set of states is referred to as Fermi sea, and the Fermi momentum $k_F$ is simply related to the nucleon  
density, $\varrho$, through  $\varrho = \nu k_F^3 / (6 \pi^2)$,  with $\nu$ being the degeneracy of the momentum eigenstates.
 The simplest implementation of the independent particle picture is the Fermi gas (FG) model, in which the mean field 
is approximated by a constant, adjusted to reproduce the values of the ground-state
energy and equilibrium density of isospin-symmetric matter inferred from nuclear data.

Despite being able to explain many properties of atomic nuclei\textemdash notably the main features of the energy spectra and the emergence of the shell 
structure\textemdash the mean-field approximation (MFA) is inherently  inadequate to take into account short-range 
correlations (SRC) among the nucleons, originating mainly from the strongly repulsive nature of nuclear forces at short distances. 
The most prominent signature of SRC is the appearance of nucleons with momenta larger than the Fermi momentum, associated with a 
substantial suppression of the probability to find two nucleons within the range of the repulsive core of the NN interaction. 
This {\it screening} effect can be accounted for by replacing the interaction terms in the Hamiltonian of Eq.~\eqref{Hamiltonian} with a
well-behaved density-dependent potential, $V_{\rm eff}$, defined through 
\begin{align}
\label{def:veff}
\langle {\overline 0} | H |  {\overline 0} \rangle =  \langle {0} |H_{\rm eff}|  {0} \rangle  = \langle {0} |~( T + V_{\rm eff} )~|  {0} \rangle 
\ \ \ \  ,  \ \  \ \ T =  \sum_{i=1}^A \frac{{\bf k}_i^2}{2 m_N} \ . 
\end{align}
Here  $| {0} \rangle$ is the FG ground state, while $|  {\overline 0} \rangle$ denotes the ground state of nuclear matter in the presence of 
correlations, defined as
\begin{align}
\label{def:F}
|  {\overline 0} \rangle = F | {0} \rangle \ \ \ \ \ , \ \ \ \ \ F  = {\color{black} {\mathcal S}} \prod_{j>i=1}^A~f_{ij} \ . 
\end{align}
The NN correlation operator $f_{ij}$ is obtained from minimization of the expectation value of $H$ in the correlated ground-state.
As a consequence, its expression,  reflecting the structure of the NN interaction, includes spin-isospin dependent and 
non central contributions. Note that, because, in general,  $[f_{ik},f_{jk}] \neq 0$, in order to preserve the symmetry of the many-body ground state
the product appearing in Eq.~\eqref{def:F} needs to be {\color{black} symmetrised  through the action of the operator ${\mathcal S}$}.
Unlike the bare $H$, the {\it renormalised} effective Hamiltonian $H_{\rm eff}$ defined by Eqs~\eqref{def:veff} and \eqref{def:F} enables perturbative calculations
of nuclear matter properties in the basis of eigenstates of the non interacting system~\cite{BF:NM,BL:2017}. 

The approach based on the formalism of Correlated Basis Functions (CBF) and the cluster expansion 
technique~\cite{CW_T,CLARK197989}\textemdash widely employed to study a variety of correlated quantum many-particle systems, ranging from atomic nuclei to nuclear matter and liquid helium~\cite{Feenberg}\textemdash
provides a consistent framework to derive the effective nuclear Hamiltonian from state-of-the-art models of the NN and NNN 
potentials~\cite{BL:2017}. The results of  these
studies have provided the input needed to carry out detailed investigations of the nuclear matter response to weak interactions based on 
realistic models of nuclear dynamics~\cite{cowell2003,cowell2004,cowell2006,benharfarina2009,Lovato:2012ux,lovatoetal2014}, which 
contributed significantly to shed light on the role of mean-field and correlation dynamics.
\begin{figure}[h!]
	\centering
	 \includegraphics[width=.475\textwidth]{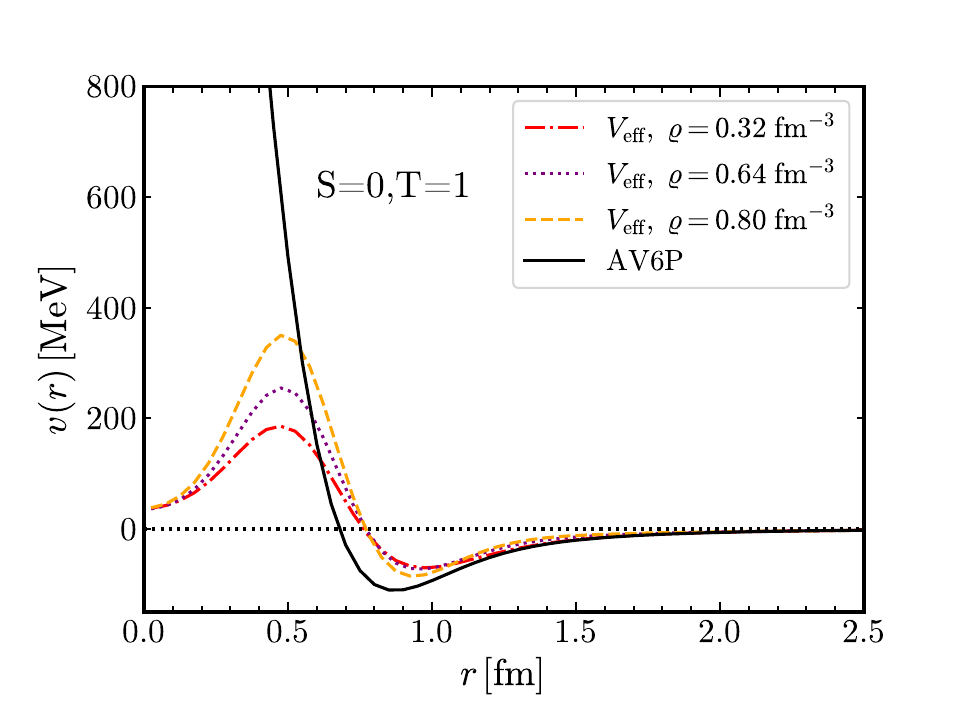}
	\caption{{\scriptsize Radial and density dependence of the effective NN interaction in the channel  of total spin-isospin $(S,T) = (0,1)$,  obtained from the authors of Ref.~\cite{BL:2017} using the CBF formalism and the cluster expansion 
technique. For comparison, the thick solid line shows the behaviour of the bare NN potential of Ref.~\cite{V6P}.  Reprinted from 
Ref.~\cite{PhysRevC.110.055801} with permissions, \copyright~APS 2024. All rights reserved.}}
	\label{sec2:fig2}
	\end{figure}

Figure~\ref{sec2:fig2} shows the radial dependence of the renormalised NN potential $V_{\rm eff}$, described in  Ref.~\cite{BL:2017},  in the 
spin-isospin channel $(S,T) = (0,1)$. The results corresponding to densities in the range relevant to neutron stars are
compared to the bare NN potential known as Argonne~$v_6^\prime$ model~\cite{V6P}. The suppression of the repulsive core 
due to NN correlations is clearly apparent.
 
 \subsection{Nuclear matter response to weak interactions}
 \label{responses}
 
 The formalism described in the previous sections, providing a consistent description of nuclear matter equilibrium properties and 
 response functions, allows to pin down the role of mean-field and correlations in a largely model-independent fashion. {\color{black} For 
 simplicity, in this Chapter we will only consider isospin-symmetric nuclear matter and pure neutron matter, which can be treated as  
 one-component Fermi systems. The extension of the formalism to the case of  $\beta$-stable matter consisting of nucleons and leptons
 does not involve any conceptual issues.}
 
 \subsubsection{Mean-field approximation}
 \label{response:MFA}
 
 Within the MFA, the final {\color{black} states $|F\rangle$ contributing to the right hand side of Eq.~\eqref{nuclear:tensor}  
 are} obtained from the
 FG ground state $|0\rangle$ by moving a nucleon from the one-particle state of momentum ${\bf p}$ and spin projection $s$, which will be denoted $| {\bf p}~s\rangle$,  to the state $ | {\bf p}^\prime~s^\prime \rangle$. It follows that $E_F - E_0 = e_{p^\prime} - e_{p}$\textemdash 
with  $e_p$ being the energy of a nucleon of momentum ${\bf p}$\textemdash and 
 the sum over final states reduces to a sum over $s$, $s^\prime$, ${\bf p}$ and ${\bf p}^\prime$. 
 Moreover, momentum conservation and Fermi-Dirac statistics 
 entail  the  constraints $|{\bf p}| < k_F$ and  $|{\bf p}^\prime| =  |{\bf p} + {\bf q}| > k_F$.

 The resulting expression of $S^{\alpha\beta}$, obtained from Eq.\eqref{nuclear:tensor}, can be written in terms of matrix elements of the 
 current $j^\alpha_i$ between single-nucleon states according to 
 \begin{align}
 \label{nuclear:response}
S^{\alpha \beta} & = \frac{1}{A} \frac{1}{2} \sum_{s s^\prime} \sum_{{\bf p}}  \langle {\bf p}~s | {j^\alpha}^\dagger | {\bf p}+{\bf q}~s^\prime \rangle
\langle {\bf p}+{\bf q}~s^\prime | {j^\beta} | {\bf p}~s \rangle 
   \ \theta(k_F - |{\bf p}|)~\big[1 - \theta(k_F - |{\bf p} + {\bf q}| ) \big]~(2 \pi)~\delta \big( \omega  + e_{p} - {e_{| {\bf p} + {\bf q}| }} \big) \ ,
 \end{align}
 where $\theta(x)$ denotes the Heaviside step function. The energy $e_p$ is often obtained within the Hartree-Fock (HF) approximation, which amounts to setting
 \begin{align}
 \label{spectrum}
 e_p = \frac{{\bf p}^2}{2m_N} + U_p + \delta e_p \ , 
 \end{align}
with $p = |{\bf p}|$ and the HF potential given by 
\begin{align}
\label{def:UHF}
U_p = \sum_{{\bf p}^\prime}~\langle {\bf p}~{\bf p}^\prime | V_{\rm eff} |  {\bf p}~{\bf p}^\prime \rangle_A~\theta \big( k_F - |{\bf p}^\prime| \big) \ , 
\end{align}
where $|  {\bf p}~{\bf p}^\prime \rangle_A = |  {\bf p}~{\bf p}^\prime \rangle - |  {\bf p}^\prime~{\bf p} \rangle$ denotes an  antisymmetrized 
two-nucleon state\footnote{ {\color{black} The dependence on the nucleon spin projection, which does not play a role in this context, is omitted 
for simplicity.} }. The additional contribution $\delta e_p$\textemdash the explicit expression of which can be found 
in, e.g., Ref.~\cite{BL:2017}\textemdash takes into account a correction originating from the density dependence of the effective interaction. The spectrum of Eq.~\eqref{spectrum} 
can be accurately approximated by  the expression
\begin{align}
\label{mstar:spectrum}
e_p = \frac{{\bf p}^2}{2 m^\star_p} + U_{p= 0}
\end{align}
with the effective mass $m^\star_p$ defined as
\begin{align}
m^\star_p  = \left[ \frac{1}{p} \left( \frac{d e_p}{d p} \right) \right]^{-1} \ .
\end{align}

The above equations define the functions $S^{\alpha \beta}$, describing the nuclear matter response to neutrino interactions
involving energy and momentum transfer $\omega$ and ${\bf q}$, respectively. The corresponding expression of the interaction 
probability, to be compared to Eq.~\eqref{W:N:NR}, reads
\begin{align}
\label{W:MFA}
W_{\rm MFA}({\bf q},\omega) = G_w^2 \Big[ S^{\varrho}(1 + \cos\theta) +  S^{\sigma} ( 3 - \cos\theta)\Big]~(2 \pi)~\delta \big( \omega  + e_{p} - {e_{| {\bf p} + {\bf q}| }} \big) \ , 
\end{align}
where the density and spin-density responses $S^{\varrho} = S^{00}$ and $S^{\sigma} = \sum_iS^{ii}/3$ are associated with the corresponding fluctuations 
induced by the so-called Fermi and Gamow-Teller transition operators {\color{black} $j^0$ and ${\bf j}$ of Eq.~\eqref{nuclear:currents} }. 

To pin down the effects of mean field nuclear dynamics, one should compare the results obtained within the MFA to the predictions of the Fermi gas model, 
in which interaction effects are neglected altogether. Note that from Eqs.~\eqref{nuclear:response} and~\eqref{W:MFA}, it follows that the
replacement of the kinetic energy spectrum with the one defined by Eqs.~\eqref{spectrum} and~\eqref{def:UHF} only affects the 
argument of the energy-conserving $\delta$-function, while the transition matrix elements of the weak current are left unchanged.

\subsubsection{Short-range correlations}
\label{response:SRC}

The results of theoretical studies provide convincing evidence that the occurrence of SRC leads to a sizable suppression of the weak 
response of nuclear matter; see, e.g., Refs.~\cite{cowell2006,benharfarina2009,Lovato:2012ux,lovatoetal2014}. This effect can be qualitatively 
understood considering that short-range NN interactions involve large momentum transfer, and therefore lead to the excitation of 
nucleons to states with momentum exceeding the Fermi momentum. On the other hand, the non vanishing occupation of states lying outside the Fermi 
sea necessarily entails a decrease of the occupation of the states with $p < k_F$ to values below unity,  which in turn reduces the 
rate of weak interactions. 

It should be emphasized that the {\it same} mechanism driving the renormalisation of the nuclear Hamiltonian is also responsible 
of the quenching of the nuclear transition amplitudes defined by Eq.\eqref{nuclear:amplitude}, which can be likewise described in terms of a 
{\it renormalised} effective operator, $j^\alpha_{\rm eff}$.
Within the approach based on correlated wave functions, the effective weak current is defined 
by the equation
\begin{align}
\label{def:jeff}
\langle {\overline F} | \sum_{i=1}^A j^\alpha_i~|  {\overline 0} \rangle = \langle {F} |~\sum_{i=1}^A j^\alpha_{\rm eff, i}~|  {0} \rangle \ , 
\end{align} 
to be compared to Eq.~\eqref{def:veff}. Here  $|  {\overline 0} \rangle$ and $|  {0} \rangle$ are the correlated and FG 
ground state of nuclear matter, respectively, while the correlated final state $|{\overline F} \rangle$ is obtained from the corresponding FG state, 
 $|F \rangle$, by applying the correlation operator of Eq.~\eqref{def:F}. The nuclear matter transition amplitudes turn out to be
\begin{align}
\label{SRC:amplitude}
\mathcal{M}^\alpha_{CBF} =  \langle F |~\sum_{i=1}^A j^\alpha_{{\rm eff},i}~| 0  \rangle  \ . 
\end{align}
The results reported in this chapter have been obtained considering only amplitudes wherein $|F \rangle$ is a 
one-particle\textendash one-hole FG state. Note, however, that  the effective current $j^\alpha_{\rm eff}$ defined by 
Eq.~\eqref{def:jeff} is a many-body operator, which may lead to transitions to more complex $n$-particle\textendash $n$-hole final states.
   
The use of a correlated wave functions also affects the form of the interaction probability, the expression of which 
can be conveniently derived in a reference frame such that the $z$ axis is in the direction of ${\bf q}$, implying $q\equiv(\omega,0,0,|{\bf q}|)$, 
with ${\bf k}$ and ${\bf k}^\prime$ lying in the $xz$ plane. The result, to be compared to Eq.~\eqref{W:MFA}, is
\begin{align}
\label{W:SRC}
W_{\rm CBF}({\bf q},\omega) = G_w^2 \Big\{~S^{\varrho}(1 + \cos\theta) 
 + S^{\sigma} (1 - \cos\theta)  + \frac{1}{2 E E^\prime} \big[ S^{11} Q_x^2  
 + S^{33}  (Q^2_z - |{\bf q}|^2) + 2~S^{13} Q_xQ_z  \big]~\Big\}
 ~(2 \pi)~\delta \big( \omega  + e_{p} - {e_{| {\bf p} + {\bf q}| }} \big) \ , 
\end{align}
where ${\bf Q} = {\bf k} + {\bf k}^\prime$, and the nuclear response functions are defined as in Eq.~\eqref{nuclear:tensor}  
{\color{black} with the amplitudes obtained from of Eqs.~\eqref{nuclear:currents} and~\eqref{SRC:amplitude}}. 

Figures~\ref{sec2:fig3}, \ref{sec2:fig4} and \ref{sec2:fig6} below illustrate the impact of nuclear dynamics on the weak response of isospin symmetric matter
at equilibrium density, $\varrho_0 = 0.16 \ {\rm fm}^{-3}$. All results correspond to charged-current Fermi interactions.

\begin{figure}[h!]
	\centering
	\includegraphics[width=.375\textwidth]{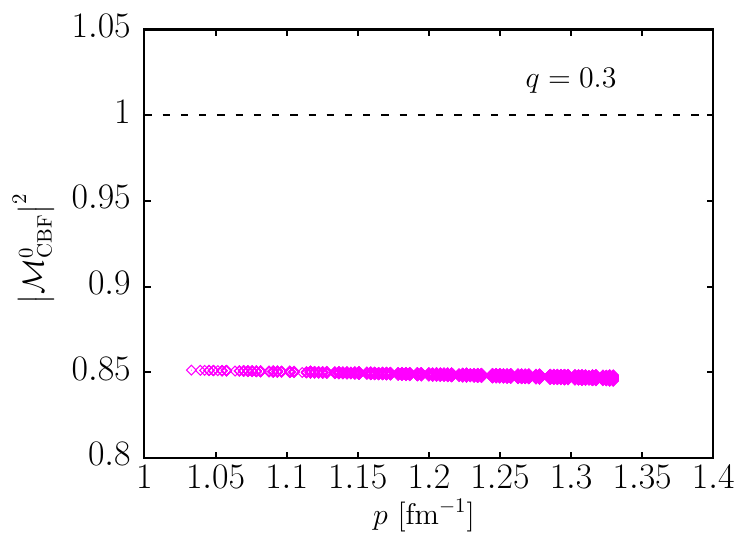} 
	\caption{{\scriptsize  Squared amplitude of the charged-current Fermi transition\textemdash defined by Eq.~\eqref{SRC:amplitude}\textemdash in isospin-symmetric 
	nuclear matter at equilibrium density, $\varrho_0 = 0.16 \ {\rm fm}^{-3}$. The results are displayed  as a function  of the 
	magnitude of the initial nucleon momentum, $p$. For comparison, 
	the dashed horizontal line shows the prediction of the FG model. Adapted from Ref.~\cite{benharfarina2009}.} }	
	\label{sec2:fig3}
	\end{figure}
 
Figure \ref{sec2:fig3} shows the squared Fermi transition amplitude $|{\mathcal M}^0_{\rm CBF}|^2$, defined by Eq.~\eqref{SRC:amplitude}, as a function of the magnitude of the initial momentum of the interacting nucleon $p$. A comparison with the prediction of the FG model, represented by the dashed horizontal line, shows that the inclusion of SRC leads to a $\sim$15\% suppression, largely independent of $p$. 
Similar results are obtained for the Gamow-Teller transition.

\begin{figure}[h!t!]
	\centering
	 \includegraphics[width=.675\textwidth]{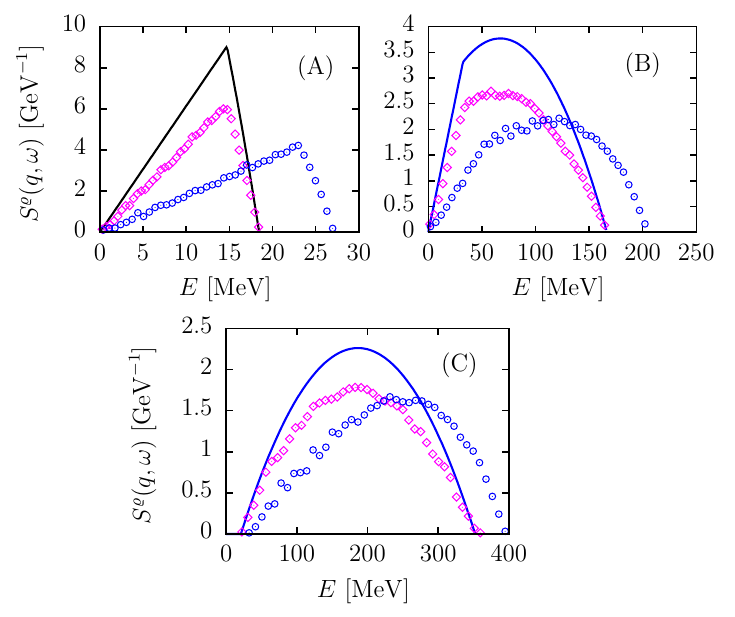}
	\caption{{\scriptsize Density response to charged-current interactions in isospin-symmetric nuclear matter at equilibrium density, corresponding to $k_F = 1.33 \  {\rm fm}^{-1}$. The solid lines shows the results of the FG model, while the squares have 
	been obtained using the CBF transition amplitude~\eqref{SRC:amplitude} and the HF spectrum of Eq.~\eqref{spectrum}. For comparison, 
	the response computed using the kinetic energy spectrum is also shown by diamonds. Panels (A), (B) and (C) correspond to 
	momentum transfer $|{\bf q}|=$0.3, 1.8 and 3.0 fm$^{-1}$, respectively. Reprinted from Ref.~\cite{benharfarina2009}. }}
	\label{sec2:fig4}
	\end{figure}

The effects of mean-field and correlation dynamics on the density response $S^\varrho$ can be observed in Fig.~\ref{sec2:fig4}, where
panels (A), (B) and (C) correspond to momentum transfer $|{\bf q}|=$0.3, 1.8 and 3.0 fm$^{-1}$, respectively. 
Note that, because the Fermi momentum of isospin-symmetric matter at equilibrium is  $k_F = 1.33 \  {\rm fm}^{-1}$, these values fall 
in the kinematic regions $|{\bf q}|<k_F$, $k_F < |{\bf q}| < 2k_F$, and  $|{\bf q}| > 2k_F$, in which the FG responses exhibit the qualitatively different behaviors clearly visible in the figure. The solid lines represent the results of the FG model, while the squares show the response
obtained from Eq.~\eqref{nuclear:response} using the CBF transition amplitudes of
Eq.~\eqref{SRC:amplitude} and the HF energy spectrum defined by of Eqs.(\ref{spectrum}) and~\eqref{def:UHF}. In the literature, 
this computational scheme is often referred to as Correlated Hartee-Fock (CHF) approximation. In order to clearly identify
the suppression due to SRC, the response obtained by using {\color{black} the} kinetic energy spectrum\textemdash that is, setting 
$U_p = \delta e_p = 0$ in Eq.~\eqref{spectrum}\textemdash is also shown by diamonds. It is apparent that, in addition to SRC, interaction effects
described by the  MFA also play a significant role, pushing strength
to energies well beyond the kinematic limit of the FG model.

\subsubsection{Long-range correlations}
\label{response:LRC}
\begin{figure}[b!]
	\centering
	 \includegraphics[width=.40\textwidth]{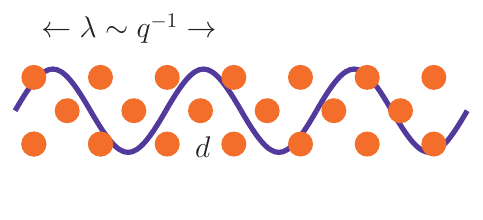}
	\caption{{\scriptsize Schematic representation of the kinematic regime in which the space resolution of 
	the beam particle,  $\lambda \sim |{\bf q}|^{-1}$, largely exceeds the NN separation distance in {\color{black} nuclear matter}.}}
	\label{sec2:fig5}
	\end{figure}

The discussion of neutrino-nucleus scattering of the previous sections was based on the unspoken premise that the beam-target 
interaction amounts to the transfer of momentum ${\bf q}$ and energy $\omega$ to a single nucleon, with the remaining particles acting 
as spectators. However, it should be kept in mind that, in order for this reaction mechanism to be dominant, the space
resolution of the incoming particle, $\lambda \sim |{\bf q}|^{-1}$, must be much shorter than the average NN separation distance 
in the target ground state. When this condition is not met, more complex mechanism, in which the momentum transfer is shared 
between many nucleons, set in, and the scattering process leaves the target in a collectively excited  state. 
The formalism outlined in the previous sections, providing a unified framework for the treatment of mean-field and correlation dynamics, 
has been also extended to describe this regime, schematically represented in Fig.~\ref{sec2:fig5}; see Refs~\cite{cowell2004,cowell2006,benharfarina2009,Lovato:2012ux,lovatoetal2014}.

It should be {\color{black} pointed out} that the FG $n$-particle\textendash $n$-hole states, while being eigenstates of the MFA 
Hamiltonian
\begin{align}
H_{MFA} = \sum_{i=1}^A h_i \ , 
\end{align}
with $h_i$ given by Eq.~\eqref{sp:h}, are {\it not} eigenstates of the full nuclear Hamiltonian. 
As a consequence, there exists a residual interaction $V_{\rm res}$ capable to induce transitions between different FG states, provided 
their momentum {\bf q}, spin and isospin are conserved.
The effects of these transitions on the weak response functions has been widely studied within the framework of the
Tamm-Dancoff approximation, also referred to as ring approximation, 
 which amounts to expanding the final state $|F\rangle$ in the basis of one-particle\textendash one-hole FG states. The corresponding energy 
 is obtained by {\color{black} numerically} solving the eigenvalue equation of the Hamiltonian $H_{MFA} + V_{\rm res}$; see, e.g., Ref.~\cite{Lovato:2012ux}.
 
\begin{figure}[h!]
	\centering
	 \includegraphics[width=.675\textwidth]{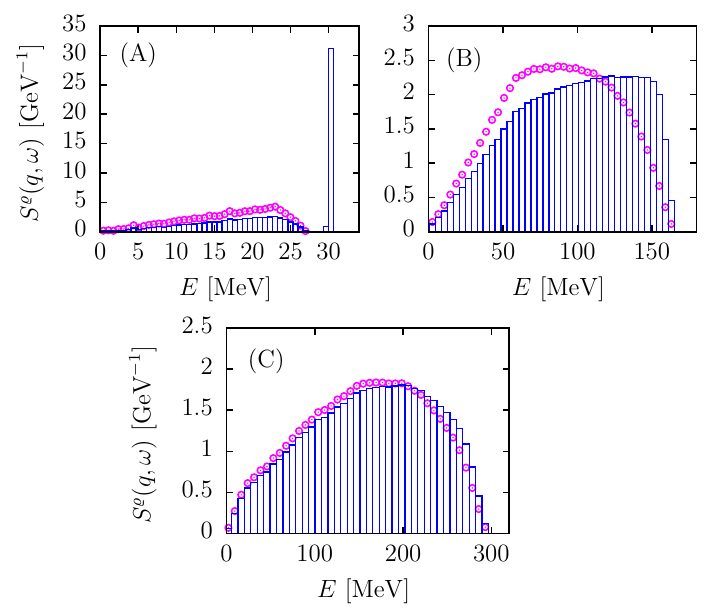}
	\caption{{\scriptsize Density response to charged-current interactions in isospin-symmetric nuclear matter at equilibrium density.
	{\color{black} Open circles and histograms} represent the results obtained using the CHF and CTD approximations described in the text. 
	Panels (A), (B) and (C) 
	correspond to momentum transfer $|{\bf q}|=$ 0,3, 1.5 and 2.4 fm$^{-1}$, respectively. Adapted from Ref.~\cite{benharfarina2009}.}}
	\label{sec2:fig6}
	\end{figure}

Figure~\ref{sec2:fig6} illustrates the results of theoretical calculations of the density 
response to charged-current interactions in isospin-symmetric nuclear 
matter at equilibrium density, computed within the Tamm-Dancoff approximation using the residual interaction obtained from 
the CBF effective Hamiltonian and the effective current operators of Eq.~\eqref{def:jeff}. This computational scheme will 
be identified as Correlated Tamm-Dancoff  (CTD) approximation. Panels (A), (B), and (C) correspond to momentum transfer $|{\bf q}|$ = 0.3, 
1.5, and 2.4 fm$^{-1}$.

The results of panel (A) show that the appearance of {\color{black} the} eigenvalue corresponding to a collective excitation,  
often referred to as {\it zero sound}, leads to the appearance of a sharp peak, clearly visible at the lowest momentum 
transfer. 
The transition to the regime in which SRC dominate is
illustrated in panels (B) and (C), showing the comparison between the 
responses at $|{\bf q}|=$ 1.5 and 2.4 fm$^{-1}$ obtained from the CHF and CTD approximations.

At $|{\bf q}|=$1.5 fm$^{-1}$ the sharp peak no longer 
sticks out, although the effect of writing the state $|F\rangle$ as a superposition of  one-particle\textendash one-hole FG states 
is still visible as an enhancement of the strength at large $\omega$. Finally, at $|{\bf q}|=$2.4 fm$^{-1}$ the role of LRC
turns out to be negligible, and the CTD and CHF responses are very close to
one another. The results of calculations of the responses associated
with Gamow-Teller transitions show a similar pattern.

\section{Neutrino mean free path in nuclear matter}
\label{response:MFP}

The mean free path (MFP) of a neutrino moving through nuclear matter with energy $E$\textemdash 
that is, the average distance $\lambda(E)$ traveled by the neutrino before colliding with a nucleon\textemdash depends on  
the number density of target particles, $\varrho$, and the total neutrino-nucleon cross section, $\sigma(E)$, according to 
$\lambda(E)^{-1} \propto \varrho~\sigma(E)$. Note that the cross section includes all contributions, arising from both scattering and absorption processes, associated with neutral- and charged-current interactions, respectively. 

The straightforward relation between cross section and interaction probability, discussed in the previous sections, allows 
one to write the neutrino MFP in the form~\cite{IW}
\begin{align} 
\label{def:lambda}
\frac{1}{\lambda(E)} = \varrho~\sigma(E) = \varrho \int \frac{d^3 k^\prime}{(2 \pi)^3} f(k^\prime) \ W({\bf q},\omega) \ , 
\end{align}
where the distribution function $f(k^\prime)$ takes into account the effect of Pauli's exclusion principle, preventing the 
appearance of {\color{black} degenerate} final-state leptons in occupied states within the Fermi sea. In the case of CC interactions 
$f(k^\prime) =  1 - \theta( {k_F}_\ell - k^\prime )$, with  ${k_F}_\ell$ being the Fermi momentum of the charged lepton $\ell$. 
In NC interactions, on the other hand, the final state lepton is a neutrino which, depending on the opacity of the nuclear medium, behaves as a degenerate or non-degenerate fermion. In the degenerate regime, corresponding to high opacity, neutrinos are 
confined within a volume $\Omega \propto R^3$ with                                                                                                                                                                                                                                                                                                                                                                                                                                                                                                                                                                                                                                                                                                                                                                                                                                                                                                                                                                                                                                                                                                                                                                                                                                                                                                                                                                                                                                                                                                                                                                                                                                                                                                                                                                                                                                                                                                                                                                                                                                                                                                                                                                                                                                                                                                                                                                                                                                                                                                                                                                                                                                                                                                                                                                                                                                                                                                                                                                                                                                                                                                                                                                                                                                                                                                                                                                                                                     finite density $\propto \Omega^{-1}$. Therefore, they have a non vanishing Fermi momentum, and obey Fermi-Dirac statistics.  
In the non-degenerate regime, on the other hand, nuclear matter is 
transparent to neutrinos, and their density vanishes, implying ${k_F}_\nu = 0$ and $f(k^\prime) \equiv 1$. 
In neutron stars, for example, the degeneracy regime sets in as soon as the neutrino MFP\textemdash which turns out to be a decreasing function of both density and temperature\textemdash exceeds the star radius, typically $\sim 10$ Km. 

Following the seminal work of Iwamoto and Pethick~\cite{IW}, the MFP of non-degenerate neutrinos corresponding to NC neutrino scattering in neutron matter has been widely studied using theoretical approaches based on the formalism of nuclear many-body theory  and realistic Hamiltonians~\cite{BCL,lovatoetal2014,Isaac:MFP}. 

\begin{figure}[h!]
	\centering
	 \includegraphics[width=.40\textwidth]{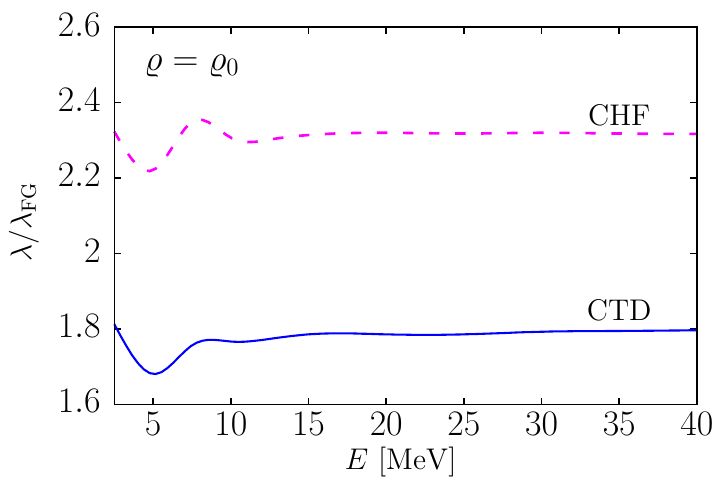}\includegraphics[width=.40\textwidth]{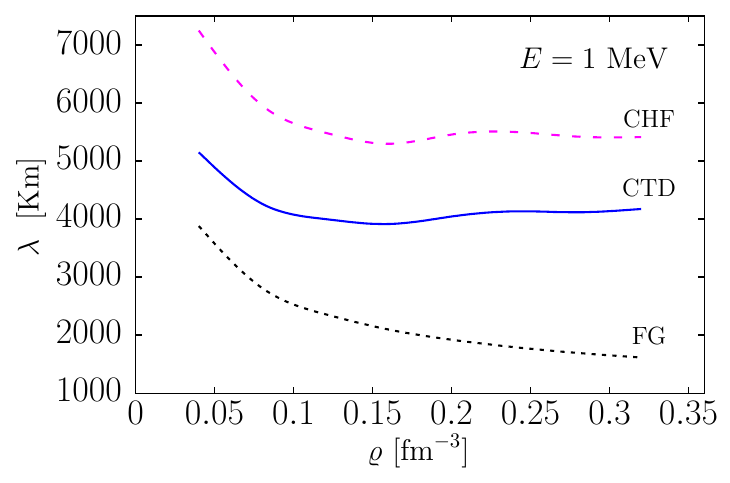}
	\caption{{\scriptsize Neutrino MFP associated with NC interactions of non-degenerate {\color{black} neutrinos} in neutron matter. Left panel: energy dependence of 
	the ratio between the MFPs at density $\varrho = \varrho_0$ reported in Ref.~\cite{lovatoetal2014} and the prediction of the FG model. 
	The dashed and solid lines represent results obtained within the  CHF and CTD approximations, respectively. Adapted from   Ref.~\cite{lovatoetal2014}.  Right panel: MFP obtained within the CHF and CTD schemes using the formalism of Landau's Fermi liquid theory and Landau parameters derived from the CBF effective interaction. Adapted from Ref.~\cite{BCL}.}}
	\label{sec2:fig7}
	\end{figure}

The results of the work of Lovato {\it et al.}~\cite{lovatoetal2014}\textemdash in which the neutrino interaction rate appearing in Eq.~\eqref{def:lambda} was obtained using the CBF effective interaction and weak current discussed in the previous sections\textemdash are illustrated in the left panel of Fig.~\ref{sec2:fig7}. 
 The ratio between {\color{black} the neutrino MFP in neutron matter} computed within the CHF approximation and the FG prediction $\lambda_{\rm FG}$, represented by the dashed line, is displayed as a function of the neutrino energy $E$ for matter density $\varrho = \varrho_0$. 
It is apparent that inclusion of the effects taken into account within  the CHF approximation  
strongly enhances the neutrino MFP.  The value of the ratio $\lambda / \lambda_{\rm FG}$ turns out to exceed 2.2 
over the whole energy range, being nearly constant  for $E\gtrsim 10$ MeV.
A comparison with the solid line, obtained using the interaction probability computed within CTD approximation shows that
the excitation of the collective zero-sound mode leads to a $\sim 25\%$ reduction..

The right panel of Fig.~\ref{sec2:fig7} illustrates the density dependence of the MFP of a 1 MeV neutrino {\color{black} in neutron matter}. 
The theoretical results of Ref.~\cite{BCL}, labelled CHF and CTD, have been obtained from the approach based on Landau's theory of normal Fermi liquid~\cite{Landau}, also used in the pioneering study of Iwamoto and Pethick~\cite{IW}. Nevertheless, they 
can be meaningfully compared to results of the left panel, because the values of the Landau parameters entering the determination of the neutrino MFP were obtained from matrix elements of the same CBF effective interaction employed in 
Ref.~\cite{lovatoetal2014}. For comparison, the Fermi gas result is also shown by the dotted line. It appears that, 
while the MFP predicted by the FG model is a monotonically decreasing function of density, the CHF and CTD results 
tend to {\color{black} become} nearly constant at densities above the equilibrium density of isospin-symmetric matter.

Overall, the picture emerging from Fig.~\ref{sec2:fig7} conforms to the expectation that the dynamical effects giving rise to the
pronounced quenching of the weak response of nuclear matter discussed in Section~\ref{nuN:inmatter}, lead to a 
corresponding sharp enhancement of the neutrino MFP.

\section{Conclusions}
\label{sec:conclusions}

The availability of a reliable theoretical framework suitable to describe the interactions of low-energy neutrinos 
with nuclear matter is the prerequisite for a quantitive understanding of a variety of astrophysical processes. 
To fully appreciate this point, just consider that the rates of the simplest processes leading to neutrino absorption and antineutrino emission in nuclear matter  
\begin{align}
\nu + n \to p + \ell   \ \ \ \ \ \ \ \ , \ \ \ \ \ \ \ \ n \to p +  {\bar \nu} + \ell \  ,
\end{align}
involve the {\it same} weak transition amplitude, that was discussed in Section~\ref{nuN:inmatter} of this chapter. 

Neutrino absorption and emission play a key role in neutron star evolution. In the early stages, 
corresponding to higher nucleon densities, neutrinos are trapped, that is, degenerate, because their MFP turns out to be much smaller than 
the stellar radius, $R \sim$10 Km. At lower densities, on the other hand, neutrinos become non degenerate, and freely escape the star carrying away energy. At this stage, neutrino emission is, in fact, the primary source of neutron star cooling. 

It should be pointed out  that, in addition to driving the thermal evolution, neutrino degeneracy also  affects the 
composition of matter in the star interior, consisting of a charge-neutral fluid of neutrons, protons and leptons in weak equilibrium. Matter composition largely determines the equation of state describing the density dependence of pressure, which in turn dictates the equilibrium properties of stable stars through the equations of Tolman, Oppenheimer and Volkoff~\cite{PhysRev.55.364,PhysRev.55.374}; 
for a concise pedagogical introduction to neutron star physics see, e.g., Ref.~\cite{benhar:LNP}.

The discussion above  is based on the premise that in dense nuclear matter temperature effects are negligible, because thermal energies are  much smaller than the typical nucleon Fermi energies. This amounts to assuming that nuclear matter is fully 
or nearly degenerate,  and can be described using the equation of state at  temperature $T=0$. 
However, thermal effects also impact the absorption and emission rates through the Fermi distributions determining the phase-space available to final-state fermions~\cite{PhysRevD.106.103020,Benhar:2023mgk}. 
Low-temperature approximations, such as the Fermi surface approximation described in, e.g., Ref.~\cite{Yakovlev},  
are routinely employed to perform calculations of the neutrino absorption and emission rates in the regime of $T \sim$~10~MeV, typical of supernovae and protoneutron stars. On the other hand, the extension of these studies to neutrino processes occurring in the postmerger phase of a binary neutron star coalescence, in which the temperature can be as high as 
$\sim$ 100~MeV~\cite{Perego_2019},  poses challenges of both conceptual and computational nature. 

The generalization of the formalism of nuclear many-body theory to describe nuclear matter at $T \neq 0$\textemdash which 
appears to be feasible in the temperature regime in which baryons are believed to be the only hadronic constituents of matter~\cite{Burgio,Benhar_2022}\textemdash will be indispensable for the interpretation of the unprecedented data
provided by multimessenger neutron star observations; for a recent review, see, e.g., Ref.~\cite{Benhar:2024qcw}.    


\begin{ack}[Acknowledgments]%
I am deeply indebted to my colleagues at INFN, Sezione di Roma, and Dipartimento di Fisica, Sapienza Universit\`a di Roma,
for countless illuminating discussions on issues related to weak interactions in nuclear matter. A special mention is owed to my 
longtime friends and collaborators Stefano Fantoni and  Valeria Ferrari, whose 
influence greatly contributed to shape my understanding of neutrino interactions, nuclear matter, and neutron stars.
\end{ack}

\input{Benhar.bbl}

\end{document}

%% file: Benhar.bbl
\begin{thebibliography*}{10}
\providecommand{\bibtype}[1]{}
\providecommand{\url}[1]{{\tt #1}}
\providecommand{\urlprefix}{URL }
\expandafter\ifx\csname urlstyle\endcsname\relax
  \providecommand{\doi}[1]{doi:\discretionary{}{}{}#1}\else
  \providecommand{\doi}{doi:\discretionary{}{}{}\begingroup
  \urlstyle{rm}\Url}\fi
\providecommand{\bibinfo}[2]{#2}
\providecommand{\eprint}[2][]{\url{#2}}
\makeatletter\def\@biblabel#1{\bibinfo{label}{[#1]}}\makeatother

\bibtype{Book}%
\bibitem{RQM}
\bibinfo{author}{L. Maiani}, \bibinfo{author}{O. Benhar},
  \bibinfo{title}{{Relativistic Quantum Mechanics}}, \bibinfo{publisher}{CRC
  Press} \bibinfo{year}{2024}, \bibinfo{doi}{\doi{10.1201/9781003436263}}.

\bibtype{Book}%
\bibitem{EW}
\bibinfo{author}{L. Maiani}, \bibinfo{title}{{Electroweak Interactions}},
  \bibinfo{publisher}{CRC Press} \bibinfo{year}{2015},
  \bibinfo{doi}{\doi{10.1201/b19048}}.

\bibtype{Book}%
\bibitem{Blatt:1952ije}
\bibinfo{author}{J.~M. Blatt}, \bibinfo{author}{V.~F. Weisskopf},
  \bibinfo{title}{{Theoretical Nuclear Physics}},
  \bibinfo{publisher}{Springer}, \bibinfo{address}{New York}
  \bibinfo{year}{1952}, \bibinfo{doi}{\doi{10.1007/978-1-4612-9959-2}}.

\bibtype{Article}%
\bibitem{RevModPhys.65.817}
\bibinfo{author}{O. Benhar}, \bibinfo{author}{V.~R. Pandharipande},
  \bibinfo{author}{S.~C. Pieper}, \bibinfo{title}{Electron-scattering studies
  of correlations in nuclei}, \bibinfo{journal}{Rev. Mod. Phys.}
  \bibinfo{volume}{65} (\bibinfo{year}{1993}) \bibinfo{pages}{817},
  \bibinfo{doi}{\doi{10.1103/RevModPhys.65.817}}.

\bibtype{Article}%
\bibitem{burrows1998}
\bibinfo{author}{A. Burrows}, \bibinfo{author}{R.~F. Sawyer},
  \bibinfo{title}{Effects of correlations on neutrino opacities in nuclear
  matter}, \bibinfo{journal}{Phys. Rev. C} \bibinfo{volume}{58}
  (\bibinfo{year}{1998}) \bibinfo{pages}{554},
  \bibinfo{doi}{\doi{10.1103/PhysRevC.58.554}}.

\bibtype{Article}%
\bibitem{burrows1999}
\bibinfo{author}{A. Burrows}, \bibinfo{author}{R.~F. Sawyer},
  \bibinfo{title}{Many-body corrections to charged-current neutrino absorption
  rates in nuclear matter}, \bibinfo{journal}{Phys. Rev. C}
  \bibinfo{volume}{59} (\bibinfo{year}{1999}) \bibinfo{pages}{510},
  \bibinfo{doi}{\doi{10.1103/PhysRevC.59.510}}.

\bibtype{Article}%
\bibitem{Dellafiore:1984ht}
\bibinfo{author}{A. Dellafiore}, \bibinfo{author}{F. Lenz},
  \bibinfo{author}{F.~A. Brieva}, \bibinfo{title}{{Particle-hole calculation of
  the longitudinal response function of \isotope[12][]{C}}},
  \bibinfo{journal}{Phys. Rev. C} \bibinfo{volume}{31} (\bibinfo{year}{1985})
  \bibinfo{pages}{1088}, \bibinfo{doi}{\doi{10.1103/PhysRevC.31.1088}}.

\bibtype{Article}%
\bibitem{Bauer_2005}
\bibinfo{author}{E Bauer}, \bibinfo{author}{A Polls}, \bibinfo{author}{A
  Ramos}, \bibinfo{title}{The longitudinal and transverse nuclear responses
  within the RPA framework}, \bibinfo{journal}{Journal of Physics G: Nuclear
  and Particle Physics} \bibinfo{volume}{31} (\bibinfo{number}{5})
  (\bibinfo{year}{2005}) \bibinfo{pages}{471},
  \bibinfo{doi}{\doi{10.1088/0954-3899/31/5/016}}.

\bibtype{Article}%
\bibitem{Horowitz}
\bibinfo{author}{C.~J. Horowitz}, \bibinfo{author}{K. Wehrberger},
  \bibinfo{title}{{Neutrino neutral current interactions in hot dense matter}},
  \bibinfo{journal}{Phys. Lett. B} \bibinfo{volume}{266} (\bibinfo{year}{1991})
  \bibinfo{pages}{236--242}, \bibinfo{doi}{\doi{10.1016/0370-2693(91)91032-Q}}.

\bibtype{Book}%
\bibitem{BF:NM}
\bibinfo{author}{O. Benhar}, \bibinfo{author}{S. Fantoni},
  \bibinfo{title}{{Nuclear Matter Theory}}, \bibinfo{publisher}{CRC Press}
  \bibinfo{year}{2020}, \bibinfo{doi}{\doi{10.1201/9781351175340}}.

\bibtype{Article}%
\bibitem{BL:2017}
\bibinfo{author}{O. Benhar}, \bibinfo{author}{A. Lovato},
  \bibinfo{title}{Perturbation theory of nuclear matter with a microscopic
  effective interaction}, \bibinfo{journal}{Phys. Rev. C} \bibinfo{volume}{96}
  (\bibinfo{year}{2017}) \bibinfo{pages}{054301},
  \bibinfo{doi}{\doi{10.1103/PhysRevC.96.054301}}.

\bibtype{Article}%
\bibitem{CW_T}
\bibinfo{author}{J.~W. Clark}, \bibinfo{author}{P. Westhaus},
  \bibinfo{title}{Method of Correlated Basis Functions},
  \bibinfo{journal}{Phys. Rev.} \bibinfo{volume}{141} (\bibinfo{year}{1966})
  \bibinfo{pages}{833}, \bibinfo{doi}{\doi{10.1103/PhysRev.141.833}}.

\bibtype{Article}%
\bibitem{CLARK197989}
\bibinfo{author}{J.~W. Clark}, \bibinfo{title}{Variational theory of nuclear
  matter}, \bibinfo{journal}{Progress in Particle and Nuclear Physics}
  \bibinfo{volume}{2} (\bibinfo{year}{1979}) \bibinfo{pages}{89},
  \bibinfo{doi}{\doi{10.1016/0146-6410(79)90004-8}}.

\bibtype{Book}%
\bibitem{Feenberg}
\bibinfo{author}{E. Feenbeg}, \bibinfo{title}{{Theory of Quantum Fluids}},
  \bibinfo{publisher}{Academic Press} \bibinfo{year}{1969},
  \bibinfo{doi}{\doi{10.1016/B978-0-12-250850-9.50001-5}}.

\bibtype{Article}%
\bibitem{cowell2003}
\bibinfo{author}{S.~T. Cowell}, \bibinfo{author}{V.~R. Pandharipande},
  \bibinfo{title}{Quenching of weak interactions in nucleon matter},
  \bibinfo{journal}{Phys. Rev. C} \bibinfo{volume}{67} (\bibinfo{year}{2003})
  \bibinfo{pages}{035504}, \bibinfo{doi}{\doi{10.1103/PhysRevC.67.035504}}.

\bibtype{Article}%
\bibitem{cowell2004}
\bibinfo{author}{S.~T. Cowell}, \bibinfo{author}{V.~R. Pandharipande},
  \bibinfo{title}{Neutrino mean free paths in cold symmetric nuclear matter},
  \bibinfo{journal}{Phys. Rev.} \bibinfo{volume}{C70} (\bibinfo{year}{2004})
  \bibinfo{pages}{035801}, \bibinfo{doi}{\doi{10.1103/PhysRevC.70.035801}}.

\bibtype{Article}%
\bibitem{cowell2006}
\bibinfo{author}{S.~T Cowell}, \bibinfo{author}{V.~R. Pandharipande},
  \bibinfo{title}{Weak interactions in hot nucleon matter},
  \bibinfo{journal}{Phys. Rev. C} \bibinfo{volume}{73} (\bibinfo{year}{2006})
  \bibinfo{pages}{025801}, \bibinfo{doi}{\doi{10.1103/PhysRevC.73.025801}}.

\bibtype{Article}%
\bibitem{benharfarina2009}
\bibinfo{author}{O. Benhar}, \bibinfo{author}{N. Farina},
  \bibinfo{title}{Correlation effects on the weak response of nuclear matter},
  \bibinfo{journal}{Phys. Lett. B} \bibinfo{volume}{680} (\bibinfo{year}{2009})
  \bibinfo{pages}{305}, \bibinfo{doi}{\doi{10.1016/j.physletb.2009.08.071}}.

\bibtype{Article}%
\bibitem{Lovato:2012ux}
\bibinfo{author}{A. Lovato}, \bibinfo{author}{C. Losa}, \bibinfo{author}{O.
  Benhar}, \bibinfo{title}{{Weak response of cold symmetric nuclear matter at
  three-body cluster level}}, \bibinfo{journal}{Nucl. Phys. A}
  \bibinfo{volume}{901} (\bibinfo{year}{2013}) \bibinfo{pages}{22},
  \bibinfo{doi}{\doi{10.1016/j.nuclphysa.2013.01.029}}.

\bibtype{Article}%
\bibitem{lovatoetal2014}
\bibinfo{author}{A. Lovato}, \bibinfo{author}{O. Benhar}, \bibinfo{author}{S.
  Gandolfi}, \bibinfo{author}{C. Losa}, \bibinfo{title}{Neutral-current
  interactions of low-energy neutrinos in dense neutron matter},
  \bibinfo{journal}{Phys. Rev. C} \bibinfo{volume}{89} (\bibinfo{year}{2014})
  \bibinfo{pages}{025804}, \bibinfo{doi}{\doi{10.1103/PhysRevC.89.025804}}.

\bibtype{Article}%
\bibitem{V6P}
\bibinfo{author}{R.B. Wiringa}, \bibinfo{author}{S.C. Pieper},
  \bibinfo{title}{{Evolution of nuclear spectra with nuclear forces}},
  \bibinfo{journal}{Phys. Rev. Lett.} \bibinfo{volume}{89}
  (\bibinfo{year}{2002}) \bibinfo{pages}{182501},
  \bibinfo{doi}{\doi{10.1103/PhysRevLett.89.182501}}.

\bibtype{Article}%
\bibitem{PhysRevC.110.055801}
\bibinfo{author}{A. Sabatucci}, \bibinfo{author}{O. Benhar},
  \bibinfo{author}{A. Lovato}, \bibinfo{title}{Relativistic corrections to the
  correlated basis function effective nuclear Hamiltonian},
  \bibinfo{journal}{Phys. Rev. C} \bibinfo{volume}{110} (\bibinfo{year}{2024})
  \bibinfo{pages}{055801}, \bibinfo{doi}{\doi{10.1103/PhysRevC.110.055801}}.

\bibtype{Article}%
\bibitem{IW}
\bibinfo{author}{N. Iwamoto}, \bibinfo{author}{C.~J. Pethick},
  \bibinfo{title}{Effects of nucleon-nucleon interactions on scattering of
  neutrinos in neutron matter}, \bibinfo{journal}{Phys. Rev. D}
  \bibinfo{volume}{25} (\bibinfo{year}{1982}) \bibinfo{pages}{313},
  \bibinfo{doi}{\doi{10.1103/PhysRevD.25.313}}.

\bibtype{Article}%
\bibitem{BCL}
\bibinfo{author}{O. Benhar}, \bibinfo{author}{A. Cipollone},
  \bibinfo{author}{A. Loreti}, \bibinfo{title}{Weak response of neutron matter
  at low momentum transfer}, \bibinfo{journal}{Phys. Rev. C}
  \bibinfo{volume}{87} (\bibinfo{year}{2013}) \bibinfo{pages}{014601},
  \bibinfo{doi}{\doi{10.1103/PhysRevC.87.014601}}.

\bibtype{Article}%
\bibitem{Isaac:MFP}
\bibinfo{author}{I. Vida\~na}, \bibinfo{author}{D. Logoteta},
  \bibinfo{author}{I. Bombaci}, \bibinfo{title}{Effect of chiral nuclear forces
  on the neutrino mean free path in hot neutron matter},
  \bibinfo{journal}{Phys. Rev. C} \bibinfo{volume}{106} (\bibinfo{year}{2022})
  \bibinfo{pages}{035804}, \bibinfo{doi}{\doi{10.1103/PhysRevC.106.035804}}.

\bibtype{Book}%
\bibitem{Landau}
\bibinfo{author}{G. Baym}, \bibinfo{author}{C.~J. Pethick},
  \bibinfo{title}{Landau Fermi-Liquid Theory}, \bibinfo{publisher}{John Wiley
  \& Sons} \bibinfo{year}{1991},
  \bibinfo{doi}{\doi{https://doi.org/10.1002/9783527617159}}.

\bibtype{Article}%
\bibitem{PhysRev.55.364}
\bibinfo{author}{R.~C. Tolman}, \bibinfo{title}{Static solutions of Einstein's
  field equations for spheres of fluid}, \bibinfo{journal}{Phys. Rev.}
  \bibinfo{volume}{55} (\bibinfo{year}{1939}) \bibinfo{pages}{364},
  \bibinfo{doi}{\doi{10.1103/PhysRev.55.364}}.

\bibtype{Article}%
\bibitem{PhysRev.55.374}
\bibinfo{author}{J.~R. Oppenheimer}, \bibinfo{author}{G.~M. Volkoff},
  \bibinfo{title}{On massive neutron cores}, \bibinfo{journal}{Phys. Rev.}
  \bibinfo{volume}{55} (\bibinfo{year}{1939}) \bibinfo{pages}{374},
  \bibinfo{doi}{\doi{10.1103/PhysRev.55.374}}.

\bibtype{Book}%
\bibitem{benhar:LNP}
\bibinfo{author}{O. Benhar}, \bibinfo{title}{{Structure and Dynamics of Compact
  Stars}}, \bibinfo{publisher}{Springer} \bibinfo{year}{2023},
  \bibinfo{doi}{\doi{10.1007/978-3-031-35628-5}}.

\bibtype{Article}%
\bibitem{PhysRevD.106.103020}
\bibinfo{author}{L. Tonetto}, \bibinfo{author}{O. Benhar},
  \bibinfo{title}{Thermal effects on nuclear matter properties},
  \bibinfo{journal}{Phys. Rev. D} \bibinfo{volume}{106} (\bibinfo{year}{2022})
  \bibinfo{pages}{103020}, \bibinfo{doi}{\doi{10.1103/PhysRevD.106.103020}}.

\bibtype{Article}%
\bibitem{Benhar:2023mgk}
\bibinfo{author}{O. Benhar}, \bibinfo{author}{A. Lovato}, \bibinfo{author}{L.
  Tonetto}, \bibinfo{title}{{Properties of Hot Nuclear Matter}},
  \bibinfo{journal}{Universe} \bibinfo{volume}{9} (\bibinfo{number}{8})
  (\bibinfo{year}{2023}) \bibinfo{pages}{345},
  \bibinfo{doi}{\doi{10.3390/universe9080345}}, \eprint{2306.01351}.

\bibtype{Article}%
\bibitem{Yakovlev}
\bibinfo{author}{D.G. Yakovlev}, \bibinfo{author}{A.D. Kaminker},
  \bibinfo{author}{O.Y. Gnedin}, \bibinfo{author}{P. Haensel},
  \bibinfo{title}{Neutrino emission from neutron stars},
  \bibinfo{journal}{Physics Reports} \bibinfo{volume}{354}
  (\bibinfo{year}{2001}) \bibinfo{pages}{1},
  \bibinfo{doi}{\doi{https://doi.org/10.1016/S0370-1573(00)00131-9}}.

\bibtype{Article}%
\bibitem{Perego_2019}
\bibinfo{author}{A. Perego}, \bibinfo{author}{S. Bernuzzi}, \bibinfo{author}{D.
  Radice}, \bibinfo{title}{Thermodynamics conditions of matter in neutron star
  mergers}, \bibinfo{journal}{The European Physical Journal A}
  \bibinfo{volume}{55} (\bibinfo{number}{8}) (\bibinfo{year}{2019}),
  \bibinfo{doi}{\doi{10.1140/epja/i2019-12810-7}}.

\bibtype{Article}%
\bibitem{Burgio}
\bibinfo{author}{G.F. Burgio}, \bibinfo{author}{H.-J. Schulze},
  \bibinfo{author}{I. Vida{\~n}a}, \bibinfo{author}{J.-B. Wei},
  \bibinfo{title}{Neutron stars and the nuclear equation of state},
  \bibinfo{journal}{Progress in Particle and Nuclear Physics}
  \bibinfo{volume}{120} (\bibinfo{year}{2021}) \bibinfo{pages}{103879},
  \bibinfo{doi}{\doi{https://doi.org/10.1016/j.ppnp.2021.103879}}.

\bibtype{Article}%
\bibitem{Benhar_2022}
\bibinfo{author}{O. Benhar}, \bibinfo{author}{A. Lovato}, \bibinfo{author}{G.
  Camelio}, \bibinfo{title}{Modeling Neutron Star Matter in the Age of
  Multimessenger Astrophysics}, \bibinfo{journal}{The Astrophysical Journal}
  \bibinfo{volume}{939} (\bibinfo{number}{1}) (\bibinfo{year}{2022})
  \bibinfo{pages}{52}, \bibinfo{doi}{\doi{10.3847/1538-4357/ac8e61}}.

\bibtype{Book}%
\bibitem{Benhar:2024qcw}
\bibinfo{editor}{O. Benhar}, \bibinfo{editor}{A. Lovato}, \bibinfo{editor}{A.
  Maselli}, \bibinfo{editor}{F. Pannarale} (Eds.), \bibinfo{title}{{Nuclear
  Theory in the Age of Multimessenger Astronomy}}, \bibinfo{publisher}{CRC
  Press} \bibinfo{year}{2024}, \bibinfo{doi}{\doi{10.1201/9781003306580}}.

\end{thebibliography*}